\documentclass[a4paper,12pt]{article}

\usepackage{bm}  
\pdfoutput=1 

\usepackage{jheppub} 

\usepackage[T1]{fontenc} 
\usepackage{amsmath}
\usepackage{amsfonts}
\usepackage{amssymb}
\usepackage{epstopdf}

\newcommand{\dso}{D_{s0}^*(2317)}
\newcommand{\dsi}{D_{s1}(2460)}
\newcommand{\dsj}{D_{sJ}(2860)}

\newcommand{\panda}{$\overline{\text P}$ANDA }

\title{Production of charm-strange hadronic molecules at the LHC}

\author[a]{Feng-Kun Guo,}
\author[a,b]{Ulf-G.  Mei\ss ner,}
\author[a]{Wei Wang}
\author[a]{and Zhi Yang}

\affiliation[a]{Helmholtz-Institut f\"ur Strahlen- und Kernphysik and Bethe Center for Theoretical Physics,
\\Universit\"at Bonn, D-53115 Bonn, Germany}
\affiliation[b]{Institute for Advanced Simulation, Institut f\"ur 
Kernphysik and J\"ulich Center for Hadron Physics,
JARA-FAME and JARA-HPC, Forschungszentrum J\"ulich, D-52425 J\"ulich, Germany}

\emailAdd{fkguo@hiskp.uni-bonn.de}
\emailAdd{meissner@hiskp.uni-bonn.de}
\emailAdd{weiwang@hiskp.uni-bonn.de}
\emailAdd{zhiyang@hiskp.uni-bonn.de}

\abstract{We explore the inclusive hadroproduction of  the 
$D_{s0}^{*}(2317)$, $D_{s1}(2460)$ and $D_{sJ}(2860)$ states 
at the Large Hadron Collider under the assumption  that these hadrons are 
$S$-wave meson-meson molecules. 
In addition, the $D_{s2}(2910)$,  a predicted  bound state of the $D_{2}(2460)K$ 
system, is also discussed. We first derive a factorisation formula 
for the production rates based on effective field theory. 
Then we make  use of two MC event generators, Herwig and Pythia, to simulate 
the production of  pairs of charmed mesons and kaons. Using effective field 
theory to handle the final state interaction among the meson 
pairs,  we give  an estimate of the inclusive production 
rates for these particles at the order-of-magnitude accuracy. Our results show that
the cross sections for the $pp\to D_{s0}^{*}(2317)$ and $pp\to D_{s1}(2460)$
at the LHC 
are at $\mathcal{O}(1~\mu$b) level, while the ones for the  $pp\to 
D_{sJ}(2860)$ and the  $pp\to  D_{s2}(2910)$   are  
smaller by about one order of magnitude. Such estimates
suggest that these four exotic states could be copiously produced at the LHC. 
A study of these states at the LHC will thus provide valuable information on 
hadron spectroscopy as well as hadron interactions. }

\begin{document} 
\maketitle
\flushbottom

\section{Introduction}
\label{sec:intro}

Heavy-light mesons can provide a unique  window to study  heavy quark dynamics
and  non-perturbative  QCD  in the presence of a heavy quark. 
In recent years, a number of the charm-strange mesons have been discovered at
various experimental facilities. These observations have not only enriched the
particle zoo of the ordinary  $c\bar s$ family,  but also have  provided 
interesting candidates for exotic  states  beyond the $\bar qq$ scenario.

One of the most interesting  exotic candidates is the  $D_{s0}^{*}(2317)$ first  
discovered in 2003 by the BABAR Collaboration in the 
inclusive $D_s\pi^0$ mass distribution from $e^{+}e^{-}$ annihilation 
data~\cite{Aubert:2003fg}. Later, the CLEO Collaboration discovered the
$D_{s1}(2460)$ in the $D^*_s\pi^0$ final state~\cite{Besson:2003cp}, followed by the 
discovery of the $D_{sJ}(2860)$ by the BABAR Collaboration in the $DK$ 
channel~\cite{Aubert:2006mh}.

The heavy-light   
hadrons are conveniently classified in spin multiplets, the members of which 
are degenerate in the  heavy quark limit $m_Q \to\infty$, labeled by the value of the 
total angular momentum of the light degrees of freedom $s_\ell$. For instance, 
the pseudoscalar charmed meson $D_{s}$ and the vector charmed meson $D^*_{s}$ belong to 
the multiplet with $s_\ell^P=\frac12^-$. Given that the mass difference 
between $D_{s1}(2460)$ and $D_{s0}^*(2317)$ is not large, 
it is likely that the $D_{s0}^{*}(2317)$ and $D_{s1}(2460)$ are spin partners. 
An important feature of the above mentioned two 
charm-strange mesons is that their masses are far below the potential model 
predictions for the lowest states with the same quantum numbers, e.g. in the 
Godfrey-Isgur quark model~\cite{Godfrey:1985xj}. Various explanations were 
proposed for these states, for reviews, see 
Refs.~\cite{Swanson:2006st,Zhu:2007wz}. Of particular interest is the hadronic 
molecule assumption under which the $\dso$ and $\dsi$ are $DK$ and $D^*K$ bound 
states, 
respectively~\cite{Barnes:2003dj,
vanBeveren:2003kd,Kolomeitsev:2003ac,Guo:2006fu,Guo:2006rp}. In this picture, 
the approximate 
equality of the mass splittings $M_{\dsi}-M_{\dso}\simeq 142$ MeV and $M_{D^*}-M_D\simeq 143$ MeV, as can 
be seen in Table~\ref{tab:mass}, is a natural consequence of heavy quark spin 
symmetry~\cite{Guo:2009id}. One can also appeal to the parity 
doublet assumption~\cite{Bardeen:2003kt,Nowak:2003ra,Mehen:2005hc} to explain this equality. 
Being below the $DK$ and $D^*K$ thresholds, the 
$D_{s0}^{*}(2317)$ and $D_{s1}(2460)$  both have narrow widths, 
less than $3.8$~MeV and $3.5$ MeV, respectively~\cite{Beringer:1900zz}.  The small widths arise from 
radiative decays and isospin-violating hadronic decay modes,  due to 
either  the $\eta-\pi^0$ mixing~\cite{Cho:1994zu}
or the mass differences between the neutral and charged charmed meson and 
kaons~\cite{Faessler:2007gv,
Faessler:2007us,Lutz:2007sk,Guo:2008gp}. These decay modes have been intensively 
explored  under   different  interpretations of the internal structure, 
and it was suggested that the isospin breaking decays can be used to 
distinguish the hadronic molecule picture from the others~\cite{Faessler:2007gv,
Faessler:2007us,Lutz:2007sk,Guo:2008gp,Liu:2012zya}. So far the 
experimental resolution is not high enough to measure the width of the hadronic 
width of the order of 100~keV predicted in the hadronic molecule picture. Yet, 
the planned \panda experiment is expected to be able to perform the 
measurement~\cite{Lutz:2009ff}. However, there is support of the hadronic 
molecule picture from some lattice QCD calculations of the charmed meson--light 
meson scattering lengths~\cite{Liu:2012zya,Mohler:2013rwa}. 

It was proposed in Ref.~\cite{Guo:2011dd} that the leading order interaction 
between an excited heavy meson with a small decay width~\footnote{The excited 
state should be narrow so that is width is small in comparison with the 
inverse of the range of forces between the heavy meson and the kaon. 
Otherwise, a new large energy scale would be introduced, and the analogy to 
the ground state mesons would be spoiled.} and the kaon should be the same as 
that for the ground state charmed mesons. This is a consequence of chiral 
symmetry when the kaons are regarded as pseudo-Goldstone bosons of the 
spontaneous breaking of chiral symmetry. Analogous to the 
$\dso$ and $\dsi$ as the $DK$ and $D^*K$ bound states, more kaonic bound states 
are predicted. Especially, the properties of the predicted $D_1(2420)K$ bound 
state agree with the experimental data for the $\dsj$ very well, including the 
ratio of branching fractions $\mathcal{B}(\dsj\to DK)/\mathcal{B}(\dsj\to 
D^*K)$ which is difficult to be accounted for in other models. Its spin 
partner, a $D_2(2460)K$ bound state, was predicted to have a mass and a width 
around 2910~MeV and 10~MeV, respectively. This state has quantum numbers 
$J^P=2^-$ and can decay into the $D^*K$ and $D_s^*\eta$. It will be called 
$D_{s2}(2910)$ throughout the paper.

\begin{table}[tbp]
\centering
\label{tab:mass}
\begin{tabular}{|l|ccc|}
\hline\hline
& $D_{s0}^{*}(2317)$ & $D_{s1}(2460)$ & $D_{sJ}(2860)$ \\\hline
constituents  &$DK$ &$D^{*}K$ &$D_{1}K$ \\
mass  &$2317.8\pm 0.6$  &$2459.6\pm 0.6$  &$2863.2^{+4.0}_{-2.6}$   \\
relative energy ($E_0$)&$-45.1\pm 0.4$ &$-44.7\pm 0.4$ &$-54.8^{+2.5}_{-1.1}$\\
\hline\hline
\end{tabular}
\caption{Masses of the $D_{s0}^{*}(2317)$, $D_{s1}(2460)$ and $D_{sJ}(2860)$ taken from the
Particle Data Group~\cite{Beringer:1900zz}, 
and the relative energy $E_0$ of the exotic states to the corresponding 
thresholds (in units of MeV).}
\end{table}

So far, most of the available experimental and theoretical investigations of 
the exotic candidates in the charm-strange sector have been focused on the 
spectrum and decays or the production in the $e^+e^-$ collisions including 
$B$ meson decays. 
In order to decipher the nature of the $D_{sJ}$ states, more accurate data 
and new processes involving these states  would be equally important.
These processes  include  heavy ion 
collisions~\cite{Cho:2010db,Cho:2011ew} and  hadroproduction.  At low 
energies, the  forthcoming \panda experiment has a 
particular  focus on the  hadroproduction of  $D_{s0}^{*}(2317)$ 
state~\cite{Lange:2009zza}, as it is supposed to be able to measure  the decay 
width of the $D_{s0}^{*}(2317)$ at the 100~keV level. 
At hadron colliders with a very high collision energy like the LHC,  the 
charm quark will be abundantly produced via the  QCD processes, which may serve 
as an ideal platform for the study of the $D_{sJ}$ states.  For instance, the 
LHCb Collaboration has reported a measurement of the decays of the 
$D_{s1}^*(2710)$, which is the first radially excited state of the $D_s^*$ with 
$J^P=1^-$, and the $D_{sJ}(2860)$ into the $D^+K^0_S$ and $D^0K^+$ final 
states with small statistical uncertainties~\cite{Aaij:2012pc}.

In this paper, we will provide a theoretical  exploration of  the 
hadroproduction   of the  $D_{s0}^{*}(2317)$, $D_{s1}(2460)$, $D_{sJ}(2860)$ and 
the $D_{s2}(2910)$ at the LHC. Our approach is similar to that used in studying
the production of  the $X(3872)$~\cite{Bignamini:2009sk,Artoisenet:2009wk,
Artoisenet:2010uu} 
and other  heavy quarkonium-like states~\cite{Guo:2013ufa,Guo:2014sca}, while for different treatments on heavy quarkonium production, see Refs.~\cite{Hou:2006it,Ali:2011qi,Ali:2013xba}.
Our calculation is based on the assumption that the four states are $S$-wave 
hadron molecules consisting of a kaon and a nonstrange charmed meson. Under 
this assumption, the charmed meson--kaon pairs will be produced first and these 
$D_{sJ}$ states are then formed through the final state interaction between 
them.  We will make  use of two Monte Carlo (MC) event generators, Herwig and 
Pythia, to simulate the production of pairs of the charmed meson and the kaon. 
Then, using  effective field theory (EFT) to handle the final state 
interaction between the kaon and charmed meson,  we will  derive  an estimate 
of the production rates for these particles at the order-of-magnitude accuracy. 
As we will show in the following, the prospect to observe these states at the 
LHC from the accumulated data is rather promising and thus an experimental 
analysis  is urgently called for. 

This paper is organised as follows.  In Sec.~\ref{sec:kaonEFT}, an overview of 
the effective field theory description of the kaonic bound states will be 
presented.  Based on the MC event generators, we will derive the 
production rates of the molecular states in Sec.~\ref{sec:productionMolecule}, 
while the numerical results  are presented subsequently  in 
Sec.~\ref{sec:numericalResults}. We will summarize this work in 
Sec.~\ref{sec:summary}.

\section{Kaonic bound states in effective field theory}
\label{sec:kaonEFT}

As the isospin of these exotic states is $I=0$, the assumed molecular 
structures can be explicitly written as
\begin{equation}
|D_{sJ}\rangle=-\frac{1}{\sqrt{2}}
\left(\left|H^+K^0\right\rangle + \left|H^0K^+\right\rangle\right) , \nonumber
\end{equation}
where $H$ denotes the $D$, $D^{*}$, $D_{1}$ and $D_{2}$ charmed mesons.

All the considered bound states contain the light pseudoscalar kaon, which 
can be considered a Goldstone boson of the spontaneous chiral symmetry 
breaking from SU$(3)_L\times$SU$(3)_R$ to SU$(3)_V$. The interaction
between the Goldstone bosons and the heavy mesons can be described by the heavy 
flavor chiral perturbation theory~\cite{Burdman:1992gh,Wise:1992hn,Yan:1992gz}, 
which combines the chiral  and heavy quark symmetries in the light and 
heavy quark sectors, respectively. The leading order contact interaction 
between a Goldstone boson and a heavy meson comes from the kinetic term in the 
Lagrangian for the heavy mesons with the derivatives being chirally gauged
\begin{equation}
    \mathcal{L}_\text{kin} = \mathcal{D}_\mu H \mathcal{D}^\mu H^\dag, 
\label{eq:Lag}
\end{equation}
where $H$ denotes the heavy meson field, and the chiral gauge derivative is given by
\begin{equation}
    \mathcal{D}_\mu = \partial_\mu + \frac12 \left( u^\dag \partial_\mu u + u\, 
\partial_\mu u^\dag \right),
\end{equation}
with $U = \exp\left(i\sqrt{2}\Phi/F \right)$ and $u = \sqrt{U}$. Here,  $\Phi$ 
is a $3\times3$ matrix parametrizing the octet Goldstone bosons
\begin{equation}
\Phi = \left(
\begin{array}{ccc}
\frac{1}{\sqrt{2}}\pi^0 + \frac{1}{\sqrt{6}}\eta & \pi^+ & K^+\\
\pi^- & -\frac{1}{\sqrt{2}}\pi^0 + \frac{1}{\sqrt{6}}\eta & K^0\\
K^- & \bar{K}^0 & - \frac{2}{\sqrt{6}}\eta
\end{array}
\right),
\end{equation}
and $F$ is the pion decay constant in the chiral limit. Since we will only 
consider the leading order interaction, $F=92.2$~MeV will be used.

\begin{figure}[t]
\centering
\includegraphics[width=0.96\linewidth]{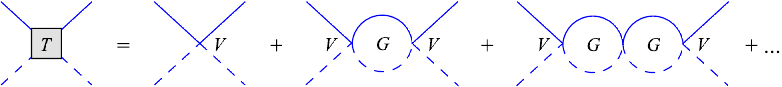}
\caption{ Sketch of the resummation of the amplitudes. Here, $T$ denotes the 
total scattering amplitude, $V$ is the leading order amplitude and $G$ is the 
loop integral. The heavy meson $H$ and kaon are represented by solid and 
dashed lines, respectively.} \label{fig:feyn}
\end{figure}
From the Lagrangian, one can derive the leading order scattering 
amplitude.~\footnote{In Ref.~\cite{Cleven:2010aw}, it is shown that the 
contribution from exchanging heavy mesons is negligible. Thus, we will only 
consider the contact terms derived from Eq.~\eqref{eq:Lag}.} One finds that 
among all the channels for the light meson--heavy meson $S$-wave scattering, 
the kaon--charmed meson one with isospin $I=0$ is the most attractive.  
The tree level scattering amplitude of the process $HK \to HK$ reads
\begin{equation}
	V(s,t, u)= - \frac{s-u}{2F^{2}}, 
	\label{eq:vstu}
\end{equation}
where $s$, $t$ and $u$ are the Mandelstam variables with the constraint $s+t+u 
= 2(m_H^2+m_K^2)$ for on-shell particles. The $S$-wave amplitude is projected 
out for on-shell particles by using
\begin{equation}
V_0(s)=\frac{1}{2}\int_{-1}^{1}V(s,t(s,\cos{\theta}),u(s,\cos{\theta}))d\cos{
\theta} = - \frac{2}{F^2} m_H E_K \left[ 1 + \mathcal{O} \left( 
\frac{k^{2}}{m_H^2} \right) \right],  
\label{eq:Vswave}
\end{equation}
where $E_K = (s-m_H^2+m_K^2)/(2\sqrt{s})$ is the energy of the heavy meson, 
$k$ is the momentum in the center-of-mass frame, and we have used the 
fact that the heavy meson is highly nonrelativistic in the  energy 
region of interest.
Since the interaction is strong, and a perturbative expansion cannot account 
for the subthreshold states such as the $\dso$, we need to sum up the 
amplitudes up to infinite orders, as shown in Fig.~\ref{fig:feyn}. This can be 
done by using the Bethe-Salpeter equation (BSE), which is an integral 
equation
\begin{equation}
      T(\bm k',\bm k; s) = V(\bm k', \bm k; s)  + i
      \int\! \frac{d^4q}{(2\pi)^4} 
\frac{ V(\bm{k}',\bm{q};s)\, T(\bm q, \bm k; 
s)  }{ (q^2-m_K^2+i\epsilon) [ (p-q)^2 -m_H^2 + i\epsilon ] },
 \label{eq:lse}
\end{equation}
where $p^2=s$, and we keep the dependence of the amplitudes on the 
center-of-mass momenta in the initial and final states explicitly.
We will follow the approach developed in Ref.~\cite{Oller:1997ti} 
and take the on-shell $S$-wave amplitude $V_0(s)$ as the kernel of the BSE.  
This is an approximation, and its justification can be found by using 
dispersion relations~\cite{Oller:1998zr,Oller:2000fj} or taking the off-shell 
contribution into account~\cite{Nieves:1999bx}.~\footnote{For a critical 
discussion on the use of the on-shell 
approximation, see e.g. Ref.~\cite{Borasoy:2007ku}.} 
Then, the integral equation is simplified to an algebra equation. The resummed 
$S$-wave amplitude reads
\begin{equation}
	T_0(s)= V_0(s) \left[ 1-G(s) V_0(s) \right]^{-1}.
    \label{eq:unitarisedAmplitudes}
\end{equation}
Notice that both here and in Eq.~\eqref{eq:vstu} we have neglected the 
polarizations of the charmed mesons with spin 1 or 2. The justification comes 
from the fact that at the energy region  of interest these charmed mesons are 
highly nonrelativistic (for details we refer to Appendix~\ref{app:pol}).
In Eq.~\eqref{eq:unitarisedAmplitudes}, $G(s)$ is the two-meson scalar loop 
function
\begin{eqnarray}
	G(s)&=& 
	 i\int \frac{d^{4}q}{(2\pi)^4}\frac{1}{q^2-m_{K}^{2}+i\epsilon} 
	 \frac{1}{(p-q)^2-m_{H}^{2}+i\epsilon}. 
\end{eqnarray}
The integral is divergent, and the divergence can be regularized using a 
subtraction constant. Treating 
the heavy meson nonrelativistically, we get an 
analytic expression for the renormalised loop function~\cite{Cleven:2010aw}
\begin{eqnarray}
G(s) =
\frac{1}{16\pi^2
  M_H}\bigg\{E_K\left[a(\mu){+}\log\left(\frac{m_K^2}{\mu^2}\right)\right]
 + 2 p_\text{cm} \cosh^{-1}\left(\frac{E_K}{m_K}\right) - i\, 2\pi
p_\text{cm} \bigg\},
\label{eq:NRloop}
\end{eqnarray}
where  $\mu$ is the regularization scale, $a(\mu)$ is the subtraction constant 
which is scale-dependent rendering the loop function scale-independent, 
$p_{cm} = \lambda(s, m_{H}^{2},m_{K}^{2})/(2\sqrt{s})$ and 
$\lambda(x,y,z)= x^{2}+y^{2}+z^{2}-2xy-2yz-2xz$ is the K\"all\'en function.  

A bound state of two mesons is associated with a pole of the resummed 
scattering amplitude below threshold on the real axis in the first Riemann 
sheet of the energy plane. Following Ref.~\cite{Guo:2011dd}, we fix the value 
of $a(\mu)$ to reproduce the mass of the $\dso$ at 2317.8~MeV, which gives 
$a(\mu=1~\text{GeV})=-3.84$, and use it in all other channels. 
Certainly, different values of the subtraction constant correspond to different
pole positions in a given channel. This allows us to vary the constant to get 
different binding energies so as to investigate the impact of binding energy on 
the cross sections later on.

\section{Hadroproduction of kaonic molecules}
\label{sec:productionMolecule}

\begin{figure}[t]
\centering
\includegraphics[width=0.42\linewidth]{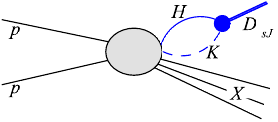}
\caption{ The mechanism considered in the paper for the inclusive production of 
the $D_{sJ}$ as a $HK$ bound state in proton--proton collisions. Here, $X$ 
denotes all the produced particles other than the $H$ and $K$ in the 
collision. } 
\label{fig:prod}
\end{figure}
In principle, the $D_{sJ}$ states can couple to other components such as the 
$c\bar s$ or a $[cq][\bar s\bar q]$ tetraquark if they have the required quantum 
numbers even the dominant component of their wave function is a kaon--charmed 
meson bound state. All 
the components contribute to the production of the $D_{sJ}$ states. In general, 
it is a process-dependent question to conclude which one is more important. 
Here we will assume that the production of the $D_{sJ}$ will occur through 
generating the charmed meson kaon pairs first which will form the $D_{sJ}$ 
states afterward. The 
mechanism is shown in Fig.~\ref{fig:prod}. This mechanism is valid only when 
the binding energy of the bound state is small so that its constituents $H$ and 
$K$ are only slightly off-shell.

\subsection{Factorization of the near-threshold production rate}
\label{sec:fact}

In Ref.~\cite{Artoisenet:2009wk}, a factorization formula for the $D\bar D^*$ 
production rate in the near-threshold region was used, which reads
\begin{eqnarray}
	d\sigma[D^{*0}\bar D^0(\bm{k})] &=& 
	 \frac{1}{\rm flux} \sum_X \int d \phi_{D^*\bar D + X} 
	 \left|\frac{{\cal M}[D^{*0}\bar D^0 (0) + X]}{T(0)}\right|^2 
|T(\bm{k})|^2 \frac{d^3k}{(2 \pi)^3 2 \mu},
	 \label{eq:DDfact}
\end{eqnarray}
where $\bm{k}$ is the center-of-mass momentum of the constituents, $\mu$ 
is the reduced mass, ${\cal M}[D^{*0}\bar D^0(\bm{k}) +X]$
is the amplitude for the $D^{*0}\bar D^0$ inclusive production, $T$ is the 
$D^{*0}\bar D^0$ scattering amplitude which accounts for the final state 
interaction (FSI) and $X$
means all the other particles in the inclusive process. The phase space 
integration runs over all the rest particles and the 
one for the total momentum of the $D^{*0}\bar D^0$ pair. In this 
FSI method, the cross-channel rescattering with other comoving particles is 
neglected. In fact, as pointed out in 
Refs.~\cite{Bignamini:2009fn,Esposito:2013ada}, the presence of such particles 
could largely affect the so-calculated cross section.

The factorization formula in Eq.~\eqref{eq:DDfact} is based on the an 
assumption that the momentum dependence of the 
production amplitude is mainly given by $T(\bm k)$ in the near-threshold 
region, and $\mathcal{M}/T$ is insensitive to $\bm k$. 
In general, however, it is not clear how $\mathcal{M}/T$ depends on the 
momentum at all. Especially, if the produced particles contain one or more of the 
lowest-lying pseudoscalar mesons (pions, kaons and the $\eta$-meson) which are 
the pseudo-Goldstone bosons of the spontaneous breaking of chiral symmetry of 
QCD, the near-threshold production amplitude without the FSI should have a 
momentum dependence dictated by chiral symmetry. At  first sight, it is not 
clear whether there exists a factorization formula analogous to 
Eq.~\eqref{eq:DDfact} or not. Here, we will show that such a formula indeed 
exists as a consequence of chiral symmetry and if we use the same on-shell 
approximation as that in Eq.~\eqref{eq:unitarisedAmplitudes}.

The general differential cross section formula for the inclusive $HK$ 
production reads
\begin{eqnarray}
	d\sigma[HK(\bm{k})] &=& 
	 \frac{1}{\rm flux} \sum_X \int d \phi_{HK + X} 
	 |{\cal M}[HK(\bm{k}) +X]|^2 \frac{d^3k}{(2 \pi)^3 2 \mu}.
	 \label{eq:generalProduction}
\end{eqnarray}
where $\bm{k}$ is the three-momentum in the center-of-mass frame of the $HK$ 
pair, $\mu$ is the reduced mass,  and ${\cal M}[HK(\bm{k}) + X]$ is the 
production amplitude.
\begin{figure}[t]
\centering
\includegraphics[width=0.8\linewidth]{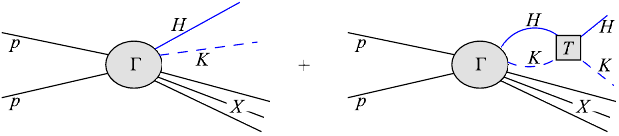}
\caption{ Inclusive production of the $HK$ pair in proton--proton collisions. 
Here $\Gamma$ denotes the direct production vertex, $T$ is the resummed $HK$
scattering amplitude, and $X$ denotes all the produced particles other than 
the $HK$ pair.} 
\label{fig:prodHK}
\end{figure}
As illustrated in Fig.~\ref{fig:prodHK}, we separate the inclusive production 
amplitude  of the $HK$ pair into two 
parts: one is due to the direct production to be denoted by $\Gamma$, and the 
other includes the FSI described by the scattering amplitude $T$. We thus 
have an integral equation
\begin{eqnarray}
    \mathcal{M}(\bm k; s) =  \Gamma(\bm k;s) + i
    \int\! \frac{d^4q}{(2\pi)^4} \frac{ \Gamma(\bm{q};s)\, T(\bm q, \bm k; s) 
}{ 
(q^2-m_K^2+i\epsilon) [ (p-q)^2 -m_H^2 + i\epsilon ] }.
\end{eqnarray}
Rewriting this equation in terms of operators, we get
\begin{equation}
    \hat{\mathcal{M}} = \hat \Gamma (1 + \hat G\, \hat T),
\end{equation}
where $\hat G$ is the operator for the Green's function of the $HK$ system. 
Similarly, the BSE given in Eq.~\eqref{eq:lse} can be written as
\begin{equation}
    \hat T = \hat V ( 1 + \hat G\, \hat T ).
\end{equation}
Therefore, we obtain
\begin{equation}
    \hat{\mathcal{M}} = \hat \Gamma \, \hat V^{-1} \, \hat T.
    \label{eq:Mfull}
\end{equation}
Since we are only considering the production of the $HK$ pair in the 
near-threshold region, the production amplitude should be dominated by the 
$S$-wave. Taking the same on-shell approximation as that in the resummed 
scattering amplitude given in Eq.~\eqref{eq:unitarisedAmplitudes}, in the 
near-threshold region we have
\begin{equation}
    \mathcal{M}(\bm k) \simeq \Gamma_0(\bm k)\, V_0(\bm k)^{-1}\, T_0(\bm k),
    \label{eq:Pswave}
\end{equation}
where we have used the center-of-mass momentum as the variable of all the 
functions.
From Eq.~\eqref{eq:Vswave} we know that $V_0(\bm k)\propto E_K = 
\sqrt{m_K^2+\bm{k}^2}$. But what is the momentum dependence of the direct 
production amplitude $\Gamma_0(\bm k)$? The answer comes from a chiral 
symmetry analysis which is applicable in the near-threshold region.

Since we are only interested in the $HK$ pair and do not care about the 
details of the other particles in the process, we may parameterize all the 
other particles involved in the inclusive production process by an external 
source field $S$ which has the same quantum numbers as the $HK$ pair. 
This means that we neglect the cross-channel 
rescattering of the charmed meson and kaon with other particles in the final 
state. The 
spirit of this kind of treatment was already used to investigate the 
$\pi\pi(K\bar K)$ system in the decays $J/\psi\to \phi\pi\pi(K\bar 
K)$~\cite{Meissner:2000bc}.  In the chiral limit when the up, down and strange 
quarks are massless, the coupling of the Goldstone bosons (pions, kaons and the 
eta) to any other fields has to be in a derivative form which ensures that the 
interaction strength vanishes at threshold as required by the Goldstone theorem. 
Thus, at leading order in the momentum expansion we have the Lagrangian for the 
near-threshold $HK$ production~\footnote{ There can be a term of the form 
$\partial^\mu S \, H\,\partial_\mu K$. However, this term can be recast into the 
one in Eq.~\eqref{eq:Lprod} modulo a higher order term by using integration by 
parts. }
\begin{equation}
    \mathcal{L}_\text{prod} = c\, S\, \partial^\mu H\, \partial_\mu K.	
\label{eq:Lprod}
\end{equation}
All the short-distance physics has been parameterized into the coefficient $c$ 
and the source field $S$. Here, we have made an implicit approximation that in 
the  final states of the inclusive production, there is no other soft chiral 
particles other than the kaon in the $HK$ pair so that we can neglect the 
interaction of the kaon with them.
Although there are two derivatives in each term, this Lagrangian is of order 
$\mathcal{O}{(k)}$, with $k$ being a momentum much smaller than the typical 
hadron scale $\Lambda_\chi\sim1$~GeV. The reason is that the dominant part of 
the heavy meson four-momentum is the mass of the heavy meson, and thus 
$\partial^\mu H$ is dominated by its temporal component. An insertion of 
the light quark masses will give higher order corrections. Therefore, at 
leading order in the 
chiral expansion, the direct production amplitude is proportional to the kaon 
energy $\Gamma_0(\bm k)\propto E_K$. The same energy dependence occurs in 
$V_0(\bm k)$ as given in Eq.~\eqref{eq:Vswave}. Together with 
Eq.~\eqref{eq:Mfull}, this means that the production amplitude can be 
factorized into the product of a constant $C$ and the $HK$ scattering amplitude 
in the near-threshold region
\begin{equation}
    \mathcal{M}(\bm k) = C \, T_0(\bm k).
\end{equation}
In brief, we have derived the factorization formula for the production of the 
$HK$ pair which is valid in the near-threshold region
\begin{eqnarray}
	d\sigma[HK(\bm{k})] &=& 
	 \frac{1}{\rm flux} \sum_X \int d \phi_{HK + X} 
	 \left| C \right|^2  
\left| T_0(\bm k) \right|^2 \frac{d^3k}{(2 \pi)^3 2 \mu},
	 \label{eq:HKProduction}
\end{eqnarray}
where the constant $C$ may take a value of $ { {\cal 
M}[HK(\bm k) + X]} / {T_0(\bm k)} $ for any  $\bm k$
provided that $k\ll \Lambda_\chi$. It is related to the coefficient $c$ in 
Eq.~\eqref{eq:Lprod} through $|C|=|c|F^2/2$.
This is the analogue of the factorization 
used in studying the $D\bar D^*$ production in Ref.~\cite{Artoisenet:2009wk}, 
and is applicable to the near-threshold production of a pair of a heavy meson 
and a pseudo-Goldstone boson. It allows for the separation of the long-distance 
and short-distance contributions in the amplitudes for the production of the 
molecules. The latter is the same for the processes $pp \to H K$ and $pp \to 
D_{sJ}$, while the long-distance factor can be deduced from the scattering 
amplitude given in Eq.~\eqref{eq:unitarisedAmplitudes}.

\subsection{Production of the charm-strange hadronic molecules}
\label{sec:proddsj}

The cross section for the $D_{sJ}$ production is
\begin{eqnarray}
	\sigma[D_{sJ}] &=& 
	 \frac{1}{\rm flux} \sum_X \int d \phi_{D_{sJ} + X} 
	 \left|{\cal M}[D_{sJ} + X] \right|^2 .
	 \label{eq:generalProductionDsJ}
\end{eqnarray}
Since in Eq.~\eqref{eq:generalProduction} the integrated phase space 
$d\phi_{HK+X}$ already contains the part of the total momentum of the $HK$ 
pair, the phase space integration in the above equation is the same as that in 
Eq.~\eqref{eq:generalProduction}.
The resummed scattering amplitude $T_0(\bm k)$ contains information about the 
generated hadronic molecules because these $D_{sJ}$ states are the bound state 
poles of $T_0(s)$. Thus, it is straightforward to extend the factorization 
formula Eq.~$\eqref{eq:HKProduction}$ to the case of the inclusive $D_{sJ}$ 
production provided that these states are produced through intermediate $HK$ 
pairs. The recipe is to replace the $HK\to HK$ scattering amplitude $T_0(s)$ in 
Eq.~\eqref{eq:HKProduction} by the amplitude for the process $HK\to D_{sJ}$, 
which is given by the effective coupling constant for the $D_{sJ}HK$ vertex in 
the vicinity of the $D_{sJ}$ pole. We have
\begin{eqnarray}
	\sigma[D_{sJ}] &=& 
	 \frac{1}{\rm flux} \sum_X \int d \phi_{D_{sJ} + X} 
	 \left| C\, g_\text{eff} \right|^2 ,
	 \label{eq:cross}
\end{eqnarray}
where $C$ is the same constant as that in Eq.~\eqref{eq:HKProduction}. 
The effective coupling constant is given by the residue of the transition 
matrix element at the pole
\begin{equation}
    |g_\text{eff}|^2 = \lim_{s\to s_\text{pole} } (s-s_\text{pole} )\, T_0(s) = 
\frac1{d[V_0(s)^{-1} - G(s)]/ds } \bigg|_{s = s_\text{pole} }. 
\end{equation}
We have checked that  the same equation will be obtained if we use 
the approach adopted in Ref.~\cite{Artoisenet:2009wk} 
which uses the Migdal-Watson theorem~\cite{Watson:1952ji,Migdal} and 
the unitary  relation for the scattering amplitude.

By varying the subtraction 
constant in the expression of loop function $G$, we can get different binding 
energies. 
Therefore the cross section for the hadronic molecule in Eq.~\eqref{eq:cross} is dependent on 
the subtraction constant, and indeed on the binding energy. With a smaller binding
energy, we found that the production rate of the molecules gets smaller. This conclusion
is in agreement with Ref~\cite{Artoisenet:2009wk}, in which the universal scattering amplitude
was adopted. From the scattering length $a=(2\mu E_B)^{-1/2}$ with $E_B$ the 
binding energy, it means that the bound
state with a larger scattering length is more difficult  to be produced.

\subsection{Estimating the cross sections using Monte Carlo event generators}

As a  phenomenological and successful  tool that has been used  in many other 
processes, MC event generators are able to simulate the 
hadronization of partons produced in  QCD processes, and therefore provide an 
 estimate of  the $pp \to HK$ inclusive cross sections. 
Two commonly adopted programs are Pythia~\cite{Sjostrand:2007gs} and Herwig~\cite{Bahr:2008pv}.
However in the above event generators, the FSI effect which governs the 
momentum dependence close to threshold, 
as depicted in the scattering amplitude $T_0(s)$ in Eq.~\eqref{eq:unitarisedAmplitudes}, 
is not incorporated. Thus, the MC cross section corresponds to the case without 
near-threshold FSI, and can be approximately expressed in terms of the vertex 
given in the leading order Lagrangian in Eq.~\eqref{eq:Lprod},
\begin{eqnarray}
	\bigg(\frac{d\sigma[HK(\bm{k})]}{dk}\bigg)_\text{MC} &=& 
	 K_{HK}\frac{1}{\rm flux} \sum_X \int d \phi_{HK + X} 
	 \left| c\, m_H\, E_K \right|^2 \frac{k^2}{4 \pi^2 \mu},
	 \label{eq:mc}
\end{eqnarray}
where $K_{HK}\sim{\cal O}(1)$ is 
introduced because of the overall difference between MC simulation and the
experimental data, while for an order-of-magnitude estimate we can roughly take 
$K_{HK}\simeq1$.

Therefore, the coupling constant $c$ can be determined using Eq.~\eqref{eq:mc}. 
Substituting the result into the Eq.~\eqref{eq:cross}, we obtain the cross 
section for the $D_{sJ}$:
\begin{eqnarray}
	\sigma[D_{sJ}] &=&  
	 \left| \frac{F^{2}}{2}\, g_\text{eff} \right|^2 
\bigg(\frac{d\sigma[HK(\bm{k})]}{dk}\bigg)_\text{MC}\frac{4\pi^{2}\mu}{k^{2}E_{K
} ^ { 2 } m_{H}^{2}}. \label{eq:cross-section-final}
\end{eqnarray}
Although the expression contains explicitly a factor of $1/(E_{K}^{2}k^2)$, this 
factor is completely cancelled by $\left( 
d\sigma[HK(\bm{k})]/dk\right)_\text{MC}$ and thus a momentum-independent value 
is obtained.

\section{Results}
\label{sec:numericalResults}

To form a molecular state, the constituents must move nearly collinear as a 
multi-quark system and thus
have a small relative momentum. Such configurations can be realized in an
inclusive $2\to 2$ QCD process, while the multiquark final states 
can be produced by soft parton shower radiations. The dominant partonic 
process for the production of the $D_{sJ}$ is $gg\to   c\bar c$, as the gluon 
density 
at  the LHC energy is much  larger than  those for quarks. 
In addition, the process $q\bar q\to c\bar c$ will  also be included in this 
analysis. 

\begin{figure*}[tbp]
\centering
\includegraphics[width=0.45\textwidth]{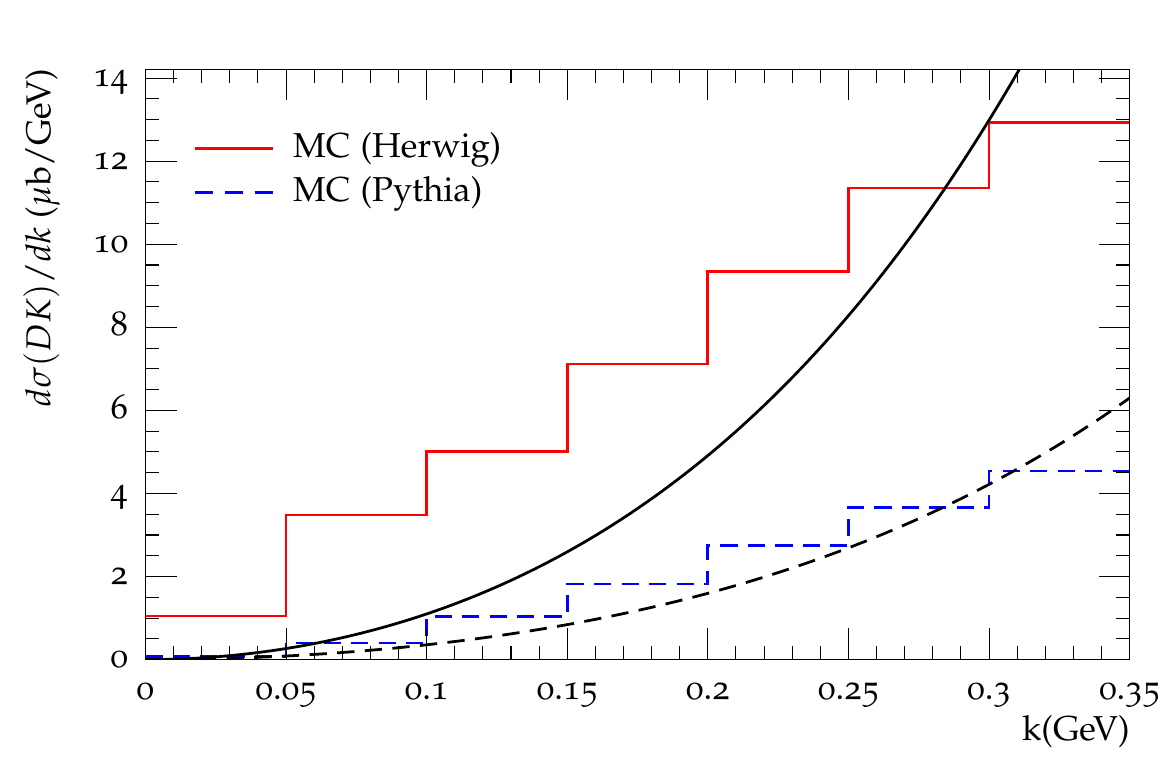}
\includegraphics[width=0.45\textwidth]{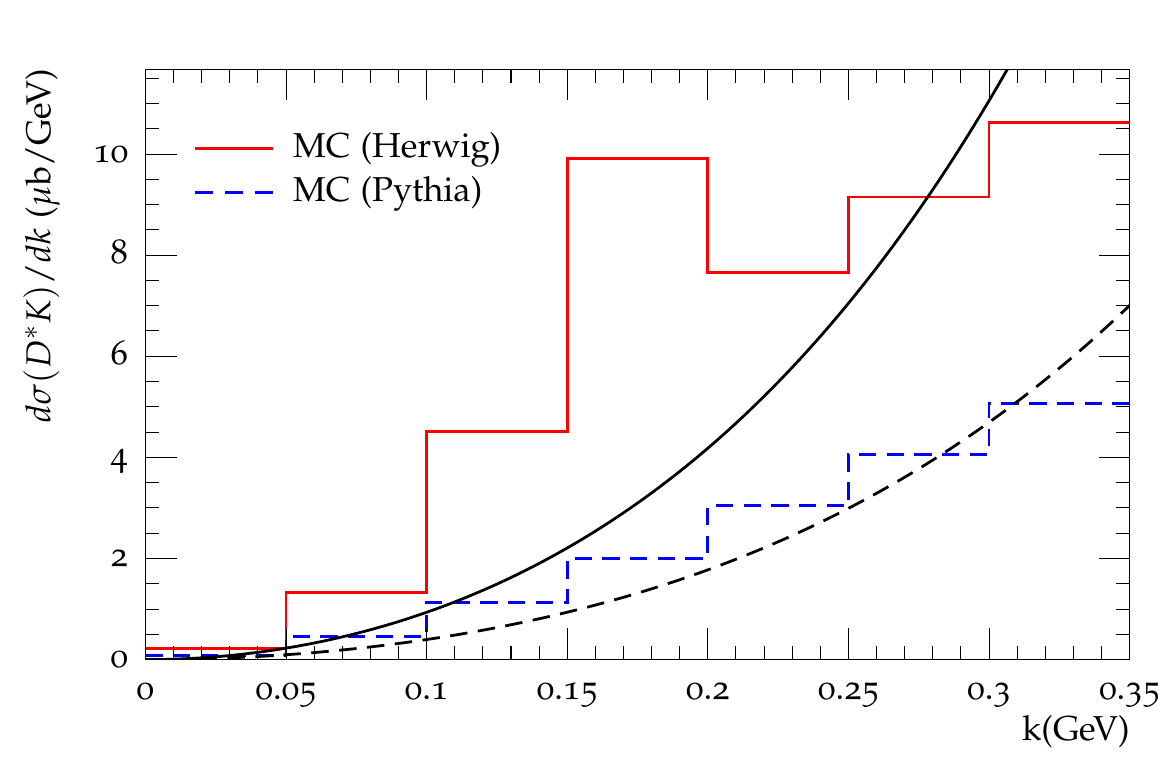} 
\includegraphics[width=0.45\textwidth]{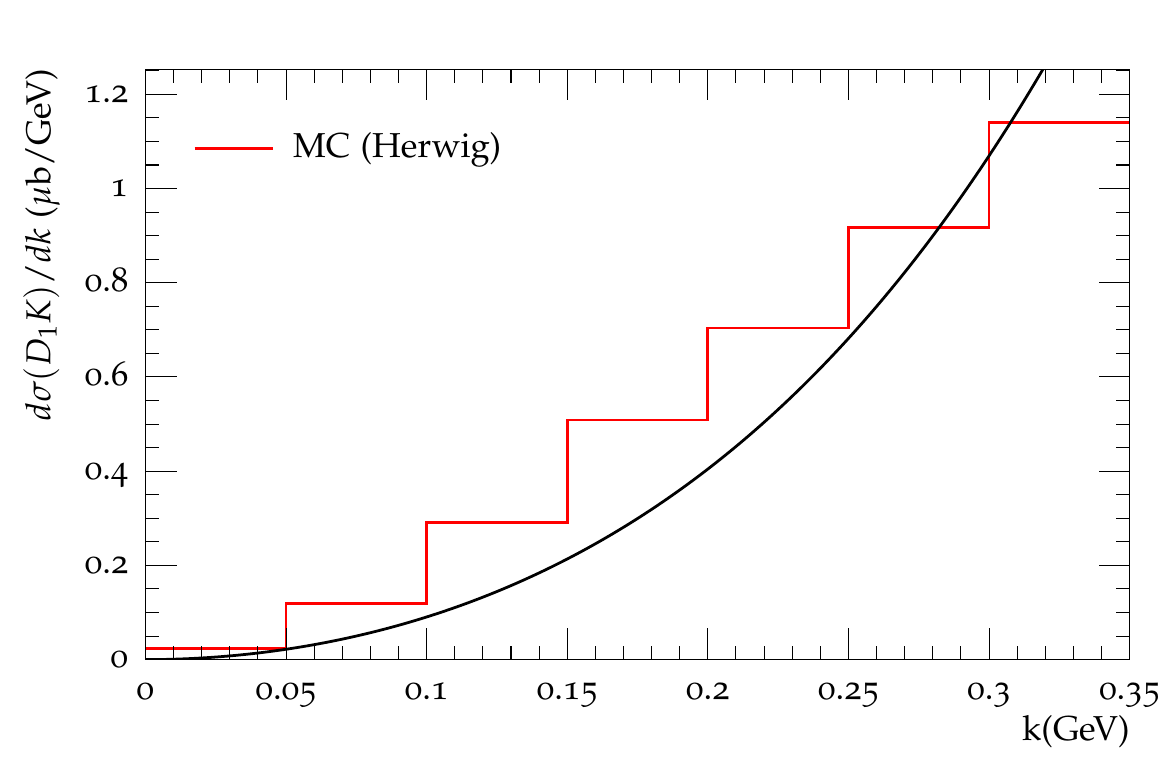}
\includegraphics[width=0.45\textwidth]{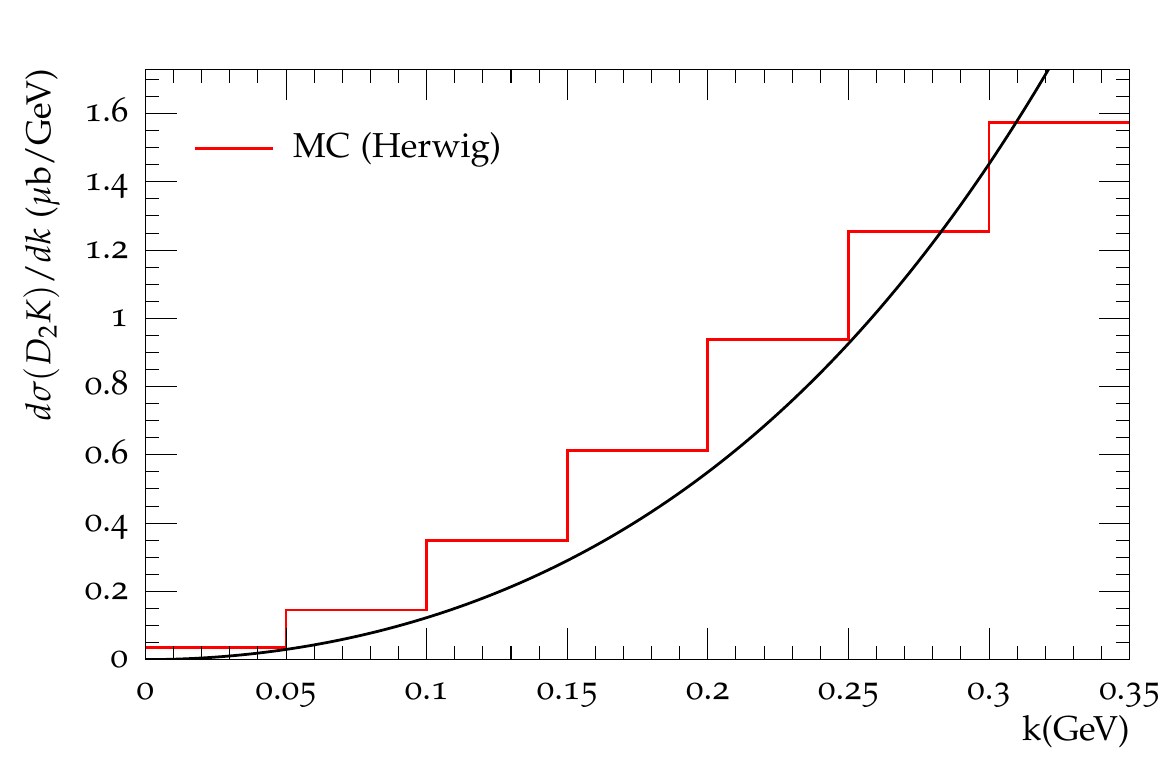}
\caption{\label{fig:diff} 
Differential cross sections ${d\sigma}/{ dk}$ (in units of $\mu$b/GeV) 
for the inclusive processes $pp\to DK$, $D^{*}K$, $D_1K$ and $D_2K$ at 
the LHC with $\sqrt s=8$ TeV. The histograms are obtained from the 
MC simulation (Herwig and Pythia) while the curves are  fitted according to the momentum dependence $k^2 E^2_K$. In the MC calculation, the kinematic cuts 
used are $|y|<2.5$ and $p_T>5$ GeV, which lie in the phase-space regions of the 
ATLAS and CMS detectors. Here, we have averaged the events in different charged 
channels, e.g. $D^0K^+$ and $D^+K^0$ for the $DK$ case.}
\end{figure*}

\begin{table}[tbp]
\centering
\begin{tabular}{|c|cccc|} \hline\hline
& $D_{s0}^{*}(2317)$ & $D_{s1}(2460)$ & $D_{sJ}(2860)$ & $D_{s2}(2910)$ \\\hline
LHC 7 &2.5(0.83) &2.1(0.91) &0.21(-) &0.27(-)\\
LHCb 7 &0.61(0.15) &0.5(0.17) &0.05(-) &0.06(-)\\
LHC 8 &2.9(0.94) &2.4(1.0) &0.24(-) &0.32(-)\\
LHCb 8 &0.74(0.18) &0.61(0.2) &0.06(-) &0.08(-)\\
LHC 14 &5.5(1.6) &4.7(1.7) &0.5(-) &0.65(-)\\
LHCb 14 &1.6(0.35) &1.3(0.38) &0.13(-) &0.17(-)\\
 \hline\hline
\end{tabular}
\caption{\label{tab:integratedCrossSection} Integrated normalized cross sections 
(in units of $\mu$b)
for the inclusive processes $pp \to 
D_{s0}^{*}(2317),\,\,D_{s1}(2460),\,\,D_{sJ}(2860)$ 
and $D_{s2}(2910)$ at LHC. The results outside (inside) brackets are 
obtained using Herwig (Pythia). Here the rapidity range $|y| <2.5$ 
has been assumed for the LHC experiments (ATLAS and CMS), 
while the rapidity range $2.0<y<4.5$ is used for the LHCb.}
\end{table}

Using Pythia and Herwig, we have generated  $10^8$ events which contain a pair 
of the charm and anti-charm quarks. These events are then analyzed by 
the Rivet 
library~\cite{Buckley:2010ar} in order  to pick out the charmed-meson kaon pair with a 
small invariant mass. Since the kaons in these events  are mostly 
produced from the soft gluon emission by the heavy charm quarks, they tend to 
move together with the charmed hadrons.  This portion of the events  will 
contribute significantly  to the formation of the molecules. 
To match the capability of the detectors, we have also implemented  the cuts on 
the transverse momentum of the meson pair $p_T>5$ GeV, and the rapidity $|y| 
<2.5$ and 
$2.0<y<4.5$   for the  ATLAS/CMS (denoted as LHC for simplicity) and LHCb  
detectors, respectively. In Fig.~\ref{fig:diff}, we show the differential cross 
sections (histograms in this figure) versus  the center-of-mass momentum  of 
the constituent mesons up to $0.35$ GeV for the inclusive processes 
$pp\to DK$, $D^{*}K$, $D_1K$, and $D_2K$ at the LHC with $\sqrt s= 8$ TeV. Since the production of
charmed mesons $D_1$ and $D_2$ is not included in Pythia, we only show the results from Herwig.

As analyzed in Sec.~\ref{sec:fact} and shown in Eq.~\eqref{eq:mc}, one should 
be able to approximately describe the production mechanism in the MC 
calculation using the leading order effective Lagrangian given in 
Eq.~\eqref{eq:Lprod}, and the resulting differential cross section is 
proportional to $k^2 E^2_K$. To validate  this feature, we fit to
these distributions with Eq.~\eqref{eq:mc} up to 350~MeV for the center-of-mass 
momentum. The fitted results  are  shown as curves in Fig.~\ref{fig:diff}. Note 
that the shape of the cross section in Eq.~\eqref{eq:mc} is completely fixed, 
and the fitting procedure only results in a normalization constant, which is 
proportional to $c$. From this figure, one sees that in most of the cases the 
MC results can be well described. The only exception is the  $D^{*}K$ from 
Herwig where there exists a clear peak at the bin between 150~MeV and 200~MeV. 
This peak can be attributed to the resonance $D_{s1}(2536)$ because it decays 
into $D^{*+}K^0$ and $D^{*0}K^+$ with a center-of-mass momentum of 149 and 
167~MeV, respectively. Note that this resonance was included in Herwig but not 
in Pythia. In principle, this resonating contribution can  be built in the 
scattering amplitudes using a coupled-channel formalism, 
which will improve the consistency with the MC simulations. However, since we are 
only interested in an order-of-magnitude estimate, we refrain from doing 
such an analysis.

With Eq.~\eqref{eq:cross-section-final} and the differential  distributions in Fig.~\ref{fig:diff}, 
we can  obtain the production rates of the hadronic $D_{sJ}$ molecules. 
These results for the cross sections 
for the inclusive processes $pp \to 
D_{s0}^{*}(2317),\,\,D_{s1}(2460),\,\,D_{sJ}(2860)$ 
and $D_{s2}(2910)$ (in units of $\mu$b) at the LHC with $\sqrt{s}= 
(7,8,14)$~TeV are shown in Table~\ref{tab:integratedCrossSection}. Results 
outside (inside) brackets are 
obtained using Herwig (Pythia if applicable). Here the rapidity range $|y| 
<2.5$ 
has been assumed for the LHC detectors (ATLAS and CMS), 
while the rapidity range $2.0<y<4.5$ is used for the LHCb. 
The differences between the values from Pythia and Herwig are caused by the 
different hadronization mechanisms  to form the charmed and kaon mesons.  From 
this table,  one sees that
the cross sections for the $pp\to D_{sJ}(2860)$ and the  $pp\to  D_{s2}(2910)$  
at the LHC are at $10^{-1}~\mu$b level,
while the ones for the $pp\to D_{s0}^{*}(2317)$ and the $pp\to D_{s1}(2460)$ 
are  larger by roughly one order of magnitude. Such large production rates  
suggest a promising perspective for searching for these
four exotic states at the LHC.

The $D_{s1}(2460)$ has a large  decay branching fraction into $D_{s}\gamma$ and thus  can be reconstructed from the
$K^+K^-\pi^+\gamma$ final states.
With the available decay branching fractions~\cite{Beringer:1900zz},
we find that the cross section for the process $pp \to D_{s1}(2460) \to D^+_s 
\gamma \to K^+K^-\pi^+\gamma$
reaches ${\cal O}$($10^7$ fb), and this would yield ${\cal O}(10^8)$ events 
when considering  the 
integrated luminosity of 22 fb$^{-1}$ from ATLAS and CMS in 
2012~\cite{ATLAS:luminosity,CMS:luminosity}.  The $D_{sJ}(2860)$ mainly decays 
into $DK$ and $D^*K$, while the $D_{s2}(2910)$ has a large partial width into 
$D^{*}K$. These two hadrons can be reconstructed in hadronic  final states.  In 
particular, based on the $1~{\rm fb}^{-1}$ data accumulated in 2011, the LHCb 
Collaboration has used the $DK$ final state to reconstruct the 
$D_{sJ}(2860)$~\cite{Aaij:2012pc}, where about $3\times 10^{4}$ $D^{0}K^{+}$ 
events from the $D_{sJ}(2860)$ were observed.  Using the cross section shown in 
Table~\ref{tab:integratedCrossSection} and assuming that the $D_{sJ}(2860)\to 
D^{0}K^{+}$ is of $\mathcal{O}(10^{-1})$ supported by the Babar 
data~\cite{Aubert:2009ah}, we predict about ${\cal O}(10^{6})$ $D^{0}K^{+}$ 
events to be generated, which is about two orders of magnitude larger. However, 
since the detection efficiency of the experiment was not published, a direct 
comparison is not possible.
Currently, we may only conclude that our prediction is not in conflict with the 
measurement. A more quantitative comparison is expected when the 
efficiency corrected data is available in future.

The search for the $D_{sJ}$ states at hadron colliders  depends on the non-resonant background 
contributions. To investigate this issue, we take the $D_{s0}(2317)$ to be 
constructed in the $D_s\gamma$ final state as an example. To be  conservative, 
we  use the cross section $\sigma(pp\to  D_s^\pm + \text{anything})$  as an 
upper bound for the 
background.
The ATLAS collaboration has provided a measurement of this cross section at 
$\sqrt s= 7$~TeV~\cite{ATLAS:2011fea}:  
\begin{eqnarray}
 \sigma(D_s^\pm)= (168\pm 34^{+27}_{-25}\pm 18 \pm 10)~{\rm \mu b}. 
\end{eqnarray}
where the  $p_T(D_s^\pm)> 3.5$ GeV and the pseudo-rapitidity $\eta< 2.1$.  Our 
results in Table~\ref{tab:integratedCrossSection} show that the cross section 
of the $pp\to  D_{s0}(2317)$ at $\sqrt s= 7$~TeV is about $2~\mu b$. Using the 
integrated luminosity in 
2012, 22~fb$^{-1}$~\cite{ATLAS:luminosity,CMS:luminosity}, we have  an estimate for the 
signal/background ratio
\begin{eqnarray}
 \frac{S}{\sqrt{B} } \sim \frac{ 2\times  22\times 10^9  
\times 5\% } { \sqrt{ 170\times  22\times 10^9   }} \sim 1\times 10^3, 
\end{eqnarray} 
where $5\%$ is the current upper bound for the branching fraction of the $D_{s0}(2317)\to D_s\gamma$~\cite{Beringer:1900zz}. 
It is worthwhile to point out that  our theoretical results have to be modified somewhat due to the mismatch in kinematics, however, 
we believe  that the above estimate  indicates a great potential for 
observing the discussed 
molecular states at the LHC.

\section{Summary}
\label{sec:summary}

Our understanding of hadron spectroscopy will  be largely improved by  studies 
of   exotic  states  that may defy the conventional 
models of $c\bar{s}$ meson spectroscopy. 
Great progress  has been made in the past decades, see Ref.~\cite{Swanson:2006st,Zhu:2007wz} 
and references therein.
One of the most important aspects is the 
discrimination of a compact multiquark configuration and a loosely bound 
hadronic molecule. Many theoretical predictions exist on the decays of the some 
of these states, and they can be used to distinguish different scenarios 
including the $c\bar s$, compact tetraquark states and hadronic molecules. 
However, only few of the decay branching fractions have been measured. 
Experimental data with high statistics are desired at this moment.

In our  work, we have  explored the hadroproduction of $D_{s0}^{*}(2317)$, 
$D_{s1}(2460)$, $D_{sJ}(2860)$ and the predicted $D_{s2}(2910)$ states 
at the LHC  under the assumption  that these hadrons are 
$S$-wave hadron molecules.  
We have made  use of two MC event generators, Herwig and Pythia, to simulate the
production of the charmed-meson kaon pairs. Together with 
effective field theory to handle the final state interaction among the meson
pairs and neglect their interactions with other particles, we 
have derived  an estimate of the production 
rates for these particles at the order-of-magnitude accuracy. Our results show that
the cross sections for the $pp\to D_{sJ}(2860)$ and the  $pp\to  D_{s2}(2910)$  
at the LHC are at $10^{-1}\mu$b level,
while the ones for the $pp\to D_{s0}^{*}(2317)$ and the $pp\to D_{s1}(2460)$ 
are  larger by roughly one order of magnitude. Thus, these states can be 
copiously produced at the LHC, and  measurements in future  would be able to 
test the molecular description of the above states  and also the production 
mechanism. Such measurements are very important to gain deeper insights into 
the hadron interactions, in particular the interactions between  heavy  
and  light mesons. 

\acknowledgments

This work is supported in part by the DFG and the NSFC 
through funds provided to the Sino-German CRC 110 ``Symmetries and the Emergence 
of Structure in QCD'', by the NSFC (Grant No. 11165005) and by the 
EU I3HP HadronPhysics3 Project.

\appendix

\section{Polarization vectors/tensors in the scattering amplitudes}
\label{app:pol}

In fact, the scattering amplitude in Eq.~\eqref{eq:vstu} has a factor of the 
product of polarization vectors (if the scattered heavy meson is the $D^*$ or 
$D_1$) or tensors (for the $D_2$), and the tree-level amplitude should be of 
the form
\begin{equation}
    V(s,t,u)\, \vec{\epsilon}^{\,(\lambda_1) } \cdot 
\vec{\epsilon}^{\,(\lambda_2) },
\end{equation}
where $\vec{\epsilon} \cdot \vec{\epsilon}$ means $\epsilon_i \epsilon_i $ if 
the scattered heavy meson is the $D^*$ or $D_1$ and $\epsilon_{ij} 
\epsilon_{ij}$ for the $D_2$, and $\lambda_{1,2}$ denote the polarizations of 
the initial and final heavy mesons. Here, we have used the fact that the heavy 
mesons are highly nonrelativistic when we are interested in the near threshold 
region, and the temporal component of polarization vector can thus be 
neglected. At one loop level of the resummed $S$-wave amplitude, one has 
\begin{equation}
    \sum_\lambda V(s) \vec\epsilon^{\,(\lambda_1) } \cdot 
\vec\epsilon^{\,(\lambda) } G(s)  V(s) \vec\epsilon^{\,(\lambda)}  \cdot 
\vec\epsilon^{\,(\lambda_2) } = \vec\epsilon^{\,(\lambda_1) } 
\cdot\vec\epsilon^{\,(\lambda_2) } V(s) G(s) V(s), \label{eq:vpol}
\end{equation}
where we have used the nonrelativistic polarization summation formulae 
\begin{eqnarray}
\sum_\lambda \epsilon_i^{\,(\lambda) }\epsilon_j^{\,(\lambda) } &=& 
\delta_{ij}, \nonumber \\
\sum_\lambda \epsilon_{ij}^{\,(\lambda) }\epsilon_{kl}^{\,(\lambda) } &=& 
\frac12 \left( \delta_{ik}\delta_{jl} + \delta_{il} \delta_{jk} \right) - 
\frac13 \delta_{ij} \delta_{kl}, 
\end{eqnarray}
and that $\epsilon_{ij}$ is symmetric and traceless. As a result of 
Eq.~\eqref{eq:vpol}, the inner product of the polarization vectors/tensors 
can be factorized out and becomes an overall factor of the resummed amplitude
\begin{equation}
\vec{\epsilon}^{\,(\lambda_1) } \cdot 
\vec{\epsilon}^{\,(\lambda_2) } \frac{V(s)}{1 - G(s) V(s) }. 
\label{eq:tpol}
\end{equation}
It looks different from the analogous equation in 
Refs.~\cite{Kolomeitsev:2003ac,Roca:2005nm}, where the denominator of the 
resummed amplitude  reads $1 
- G(s) V(s) \left[1+ q^2_\text{cm}/(3M_V)\right]$ for the pseudoscalar 
meson--vector meson scattering, where $q_\text{cm}$ is the size of the momentum 
of the vector meson in the center-of-mass frame, and $M_V$ is the vector 
meson mass. In the nonrelativistic limit, the additional factor $1+ 
q^2_\text{on}/(3M_V)$ is reduced to 1, and one gets the same equation as 
Eq.~\eqref{eq:tpol}.



\bigskip

\begin{thebibliography}{99}

\small

\bibitem{Aubert:2003fg} 
  B.~Aubert {\it et al.}  [BaBar Collaboration],
  \emph{Observation of a narrow meson decaying to $D_s^+ \pi^0$ at a mass of 
2.32~GeV/$c^2$},
  \emph{Phys. Rev. Lett.}  \textbf{90}, 242001 (2003)
  [hep-ex/0304021].

\bibitem{Besson:2003cp} 
  D.~Besson {\it et al.}  [CLEO Collaboration],
  \emph{Observation of a narrow resonance of mass 2.46 GeV$/c^2$ decaying 
to $D^{*+}_s \pi^0$ and confirmation of the $D^*_{sJ}(2317)$ state},
  \emph{Phys. Rev.} \textbf{D 68}, 032002 (2003)
  [Erratum-ibid. \textbf{D 75}, 119908 (2007)]
  [hep-ex/0305100].

\bibitem{Aubert:2006mh} 
  B.~Aubert {\it et al.}  [BaBar Collaboration],
  \emph{Observation of a New D(s) Meson Decaying to DK at a Mass of 
2.86 GeV$/c^2$},
  \emph{Phys. Rev. Lett.}  \textbf{97}, 222001 (2006)
  [hep-ex/0607082].

\bibitem{Godfrey:1985xj} 
  S.~Godfrey and N.~Isgur,
  \emph{Mesons in a Relativized Quark Model with Chromodynamics},
  \emph{Phys. Rev.} \textbf{D 32}, 189 (1985).
  
\bibitem{Swanson:2006st} 
  E.~S.~Swanson,
  \emph{The New heavy mesons: A Status report},
  \emph{Phys. Rept.}  {\bf 429}, 243 (2006)
  [hep-ph/0601110].
  
  
\bibitem{Zhu:2007wz} 
  S.-L.~Zhu,
  \emph{New hadron states},
  \emph{Int. J. Mod. Phys.} \textbf{E 17}, 283 (2008)
  [hep-ph/0703225].
  
\bibitem{Barnes:2003dj} 
  T.~Barnes, F.~E.~Close and H.~J.~Lipkin,
  \emph{Implications of a DK molecule at 2.32 GeV},
  \emph{Phys. Rev.} \textbf{D 68}, 054006 (2003)
  [hep-ph/0305025].
  
\bibitem{vanBeveren:2003kd} 
  E.~van Beveren and G.~Rupp,
  \emph{Observed $D_s(2317)$ and tentative $D(2030)$ as the charmed cousins of 
the light scalar nonet},
  \emph{Phys.\ Rev.\ Lett.}  {\bf 91}, 012003 (2003)
  [hep-ph/0305035].
  
\bibitem{Kolomeitsev:2003ac} 
  E.~E.~Kolomeitsev and M.~F.~M.~Lutz,
  \emph{On Heavy light meson resonances and chiral symmetry},
  \emph{Phys.\ Lett.} {\bf B 582}, 39 (2004)
  [hep-ph/0307133].
  
\bibitem{Guo:2006fu} 
  F.-K.~Guo, P.-N.~Shen, H.-C.~Chiang, R.-G.~Ping and B.-S.~Zou,
  \emph{Dynamically generated 0+ heavy mesons in a heavy chiral unitary 
approach},
  \emph{Phys.\ Lett.}  {\bf B 641}, 278 (2006)
  [hep-ph/0603072].
  
\bibitem{Guo:2006rp} 
  F.-K.~Guo, P.-N.~Shen and H.-C.~Chiang,
  \emph{Dynamically generated 1+ heavy mesons},
  \emph{Phys.\ Lett.} {\bf B 647}, 133 (2007)
  [hep-ph/0610008].
  
\bibitem{Guo:2009id} 
  F.-K.~Guo, C.~Hanhart and U.-G.~Mei\ss{}ner,
  \emph{Implications of heavy quark spin symmetry on heavy meson hadronic 
molecules},
  \emph{Phys.\ Rev.\ Lett.}  {\bf 102}, 242004 (2009)
  [arXiv:0904.3338 [hep-ph]].
  
\bibitem{Bardeen:2003kt} 
  W.~A.~Bardeen, E.~J.~Eichten and C.~T.~Hill,
  \emph{Chiral multiplets of heavy-light mesons},
  \emph{Phys.\ Rev.} D {\bf 68}, 054024 (2003)
  [hep-ph/0305049].
  
\bibitem{Nowak:2003ra} 
  M.~A.~Nowak, M.~Rho and I.~Zahed,
  \emph{Chiral doubling of heavy light hadrons: BABAR 2317~MeV$/c^2$ and CLEO 
2463~MeV$/c^2$ discoveries},
  \emph{Acta Phys.\ Polon.} B {\bf 35}, 2377 (2004)
  [hep-ph/0307102].
  
\bibitem{Mehen:2005hc} 
  T.~Mehen and R.~P.~Springer,
  \emph{Even- and odd-parity charmed meson masses in heavy hadron chiral 
perturbation theory},
  \emph{Phys.\ Rev.} D {\bf 72}, 034006 (2005)
  [hep-ph/0503134].


\bibitem{Beringer:1900zz} 
  J.~Beringer {\it et al.}  [Particle Data Group],
  \emph{Review of Particle Physics (RPP)},
  \emph{Phys. Rev.} \textbf{D 86}, 010001 (2012).


\bibitem{Cho:1994zu} 
  P.~L.~Cho and M.~B.~Wise,
  \emph{Comment on $D_s^* \to D_s\pi^0$ decay},
  \emph{Phys. Rev.} \textbf{D 49}, 6228 (1994)
  [hep-ph/9401301].
  

\bibitem{Faessler:2007gv} 
  A.~Faessler, T.~Gutsche, V.~E.~Lyubovitskij and Y.~-L.~Ma,
  \emph{Strong and radiative decays of the $D_{s0}^*(2317)$ meson in the 
$DK$ molecule picture},
  \emph{Phys. Rev.} \textbf{D 76}, 014005 (2007)
  [arXiv:0705.0254 [hep-ph]].
  
\bibitem{Faessler:2007us} 
  A.~Faessler, T.~Gutsche, V.~E.~Lyubovitskij and Y.~-L.~Ma,
  \emph{$D^*K$ molecular structure of the $D_{s1}(2460)$ meson},
  \emph{Phys. Rev.} \textbf{D 76}, 114008 (2007)
  [arXiv:0709.3946 [hep-ph]].
  
\bibitem{Lutz:2007sk} 
  M.~F.~M.~Lutz and M.~Soyeur,
  \emph{Radiative and isospin-violating decays of $D_s$-mesons in the 
hadrogenesis conjecture},
  \emph{Nucl. Phys.} \textbf{A 813}, 14 (2008)
  [arXiv:0710.1545 [hep-ph]].
  
\bibitem{Guo:2008gp} 
  F.-K.~Guo, C.~Hanhart, S.~Krewald and U.-G.~Mei{\ss}ner,
  \emph{Subleading contributions to the width of the $D^*_{s0}(2317)$},
  \emph{Phys. Lett.} \textbf{B 666}, 251 (2008)
  [arXiv:0806.3374 [hep-ph]].
  
\bibitem{Liu:2012zya} 
  L.~Liu, K.~Orginos, F.-K.~Guo, C.~Hanhart and U.-G.~Mei\ss{}ner,
  \emph{Interactions of Charmed Mesons with Light Pseudoscalar Mesons from Lattice QCD and Implications on the Nature of the $D_{s0}^*(2317)$},
  \emph{Phys. Rev.} \textbf{D 87}, 014508 (2013)
  [arXiv:1208.4535 [hep-lat]].

  

\bibitem{Lutz:2009ff} 
  W.~Erni {\it et al.}  [PANDA Collaboration], M.~F.~M.~Lutz, B.~Pire, 
O.~Scholten and R.~Timmermans,
  \emph{Physics Performance Report for PANDA: Strong Interaction Studies with 
Antiprotons,}
  arXiv:0903.3905 [hep-ex].
  
  


\bibitem{Mohler:2013rwa} 
  D.~Mohler, C.~B.~Lang, L.~Leskovec, S.~Prelovsek and R.~M.~Woloshyn,
  \emph{$D_{s0}^*(2317)$ Meson and $D$-Meson-Kaon Scattering from Lattice QCD},
  \emph{Phys.\ Rev.\ Lett.}  {\bf 111}, 222001 (2013)
  [arXiv:1308.3175 [hep-lat]].
  
  
\bibitem{Guo:2011dd} 
  F.-K.~Guo and U.-G.~Mei\ss{}ner,
  \emph{More kaonic bound states and a comprehensive interpretation of the $D_{sJ}$ states},
  \emph{Phys. Rev.} \textbf{D 84}, 014013 (2011)
  [arXiv:1102.3536 [hep-ph]].



\bibitem{Cho:2010db} 
  S.~Cho {\it et al.}  [ExHIC Collaboration],
  \emph{Multi-quark hadrons from Heavy Ion Collisions},
  Phys.\ Rev.\ Lett.\  {\bf 106}, 212001 (2011)
  [arXiv:1011.0852 [nucl-th]].

\bibitem{Cho:2011ew} 
  S.~Cho {\it et al.}  [ExHIC Collaboration],
  \emph{Studying Exotic Hadrons in Heavy Ion Collisions},
  Phys.\ Rev.\ C {\bf 84}, 064910 (2011)
  [arXiv:1107.1302 [nucl-th]].


\bibitem{Lange:2009zza} 
  J.~S.~Lange [PANDA Collaboration],
  \emph{The PANDA experiment: Hadron physics with antiprotons at FAIR},
  \emph{Int. J. Mod. Phys.} \textbf{A 24}, 369 (2009).


\bibitem{Aaij:2012pc} 
  R.~Aaij {\it et al.}  [LHCb Collaboration],
  \emph{Study of $D_{sJ}$ decays to $D^+K_S$ and $D^0K^+$ final states in $pp$ collisions},
  \emph{JHEP} \textbf{1210}, 151 (2012)
  [arXiv:1207.6016 [hep-ex]].
  
 
\bibitem{Bignamini:2009sk} 
  C.~Bignamini, B.~Grinstein, F.~Piccinini, A.~D.~Polosa and C.~Sabelli,
  \emph{Is the X(3872) Production Cross Section at Tevatron Compatible with a 
Hadron Molecule Interpretation?},
  \emph{Phys. Rev. Lett.}  \textbf{103}, 162001 (2009)
  [arXiv:0906.0882 [hep-ph]]. 
  
  
\bibitem{Artoisenet:2009wk} 
  P.~Artoisenet and E.~Braaten,
  \emph{Production of the X(3872) at the Tevatron and the LHC},
  \emph{Phys. Rev.} \textbf{D 81}, 114018 (2010)
  [arXiv:0911.2016 [hep-ph]].

\bibitem{Artoisenet:2010uu} 
  P.~Artoisenet and E.~Braaten,
  \emph{Estimating the Production Rate of Loosely-bound Hadronic Molecules using Event Generators},
  \emph{Phys. Rev.} \textbf{D 83}, 014019 (2011)
  [arXiv:1007.2868 [hep-ph]].

  
\bibitem{Guo:2013ufa} 
  F.-K.~Guo, U.-G.~Mei\ss{}ner and W.~Wang,
  \emph{Production of charged heavy quarkonium-like states at the LHC and the Tevatron},
  arXiv:1308.0193 [hep-ph].
  
  
\bibitem{Guo:2014sca} 
  F.~-K.~Guo, U.~-G.~Mei\ss{}ner and W.~Wang,
  \emph{Production of the bottom analogues and the spin partner of the X(3872) at hadron colliders},
  arXiv:1402.6236 [hep-ph].
  
 
\bibitem{Hou:2006it} 
  W.-S.~Hou,
  \emph{Searching for the bottom counterparts of $X(3872) $ and $Y(4260)$ via 
$\pi^{+} \pi^{-} \Upsilon$},
  Phys.\ Rev.\ D {\bf 74}, 017504 (2006)
  [hep-ph/0606016].
  
\bibitem{Ali:2011qi} 
  A.~Ali and W.~Wang,
  \emph{Production of the Exotic $1^{--}$ Hadrons $\phi(2170)$, X(4260) and $Y_b(10890)$ at the LHC and Tevatron via the Drell-Yan Mechanism},
  \emph{Phys.\ Rev.\ Lett.}  {\bf 106}, 192001 (2011)
  [arXiv:1103.4587 [hep-ph]].
  
\bibitem{Ali:2013xba} 
  A.~Ali, C.~Hambrock and W.~Wang,
  \emph{Hadroproduction of $\Upsilon(nS)$ above $B\bar B$ Thresholds and Implications for $Y_b(10890)$},
  \emph{Phys.\  Rev.}  D 88, {\bf 054026} (2013)
  [arXiv:1306.4470 [hep-ph]].

\bibitem{Burdman:1992gh} 
  G.~Burdman and J.~F.~Donoghue,
  \emph{Union of chiral and heavy quark symmetries},
  \emph{Phys. Lett.} \textbf{B 280}, 287 (1992).

\bibitem{Wise:1992hn} 
  M.~B.~Wise,
  \emph{Chiral perturbation theory for hadrons containing a heavy quark},
  \emph{Phys. Rev.} \textbf{D 45}, 2188 (1992).

\bibitem{Yan:1992gz} 
  T.~-M.~Yan, H.~-Y.~Cheng, C.~-Y.~Cheung, G.~-L.~Lin, Y.~C.~Lin and H.~-L.~Yu,
  \emph{Heavy quark symmetry and chiral dynamics},
  \emph{Phys. Rev.} \textbf{D 46}, 1148 (1992)
  [Erratum-ibid. \textbf{D 55}, 5851 (1997)].

 
\bibitem{Cleven:2010aw} 
  M.~Cleven, F.-K.~Guo, C.~Hanhart and U.-G.~Mei\ss{}ner,
  \emph{Light meson mass dependence of the positive parity heavy-strange 
mesons},
  \emph{Eur.\ Phys.\ J.} A {\bf 47}, 19 (2011)
  [arXiv:1009.3804 [hep-ph]].  
  
  
\bibitem{Bignamini:2009fn} 
  C.~Bignamini, B.~Grinstein, F.~Piccinini, A.~D.~Polosa, V.~Riquer and 
C.~Sabelli,
  \emph{More loosely bound hadron molecules at CDF?,}
  \emph{Phys.\ Lett.}  {\bf B 684}, 228 (2010)
  [arXiv:0912.5064 [hep-ph]].  

\bibitem{Esposito:2013ada} 
  AEsposito, FPiccinini, APilloni and A.~D.~Polosa,
  \emph{A Mechanism for Hadron Molecule Production in $p\bar p(p)$
Collisions,}
  \emph{J.\ Mod.\ Phys.}  {\bf 4}, 1569 (2013)
  [arXiv:1305.0527 [hep-ph]].
  
\bibitem{Oller:1997ti} 
  J.~A.~Oller and E.~Oset,
  \emph{Chiral symmetry amplitudes in the $S$ wave isoscalar and isovector 
channels 
  and the $\sigma$, $f_0(980)$, $a_0(980)$ scalar mesons},
  \emph{Nucl. Phys.} \textbf{A 620}, 438 (1997)
  [Erratum-ibid. \textbf{A 652}, 407 (1999)]
  [hep-ph/9702314].  

\bibitem{Oller:1998zr} 
  J.~A.~Oller and E.~Oset,
  \emph{N/D description of two meson amplitudes and chiral symmetry},
  \emph{Phys. Rev.} \textbf{D 60}, 074023 (1999)
  [hep-ph/9809337].

\bibitem{Oller:2000fj} 
  J.~A.~Oller and U.-G.~Mei\ss{}ner,
  \emph{Chiral dynamics in the presence of bound states: Kaon nucleon 
interactions   revisited},
  \emph{Phys.\ Lett.} {\bf B 500}, 263 (2001)
  [hep-ph/0011146].

\bibitem{Nieves:1999bx} 
  J.~Nieves and E.~Ruiz Arriola,
  \emph{Bethe-Salpeter approach for unitarized chiral perturbation theory},
  \emph{Nucl.\ Phys.} {\bf A 679}, 57 (2000)
  [hep-ph/9907469].
  
\bibitem{Borasoy:2007ku} 
  B.~Borasoy, P.~C.~Bruns, U.-G.~Mei{\ss}ner and R.~Ni{\ss}ler,
  \emph{A Gauge invariant chiral unitary framework for kaon photo- and 
electroproduction on the proton},
  \emph{Eur.\ Phys.\ J.}  {\bf A 34}, 161 (2007)
  [arXiv:0709.3181 [nucl-th]].
 
\bibitem{Meissner:2000bc} 
  U.-G.~Mei\ss{}ner and J.~A.~Oller,
  \emph{$J/\psi \to \phi \pi \pi (K \bar K)$ decays, chiral dynamics and OZI 
  violation},
  \emph{Nucl.\ Phys.} {\bf A 679}, 671 (2001)
  [hep-ph/0005253]. 
  
\bibitem{Watson:1952ji} 
  K.~M.~Watson,
  \emph{The Effect of final state interactions on reaction cross-sections},
  \emph{Phys. Rev.}  \textbf{88}, 1163 (1952).

\bibitem{Migdal}  
  A.~B.~Migdal,
  \emph{The theory of nuclear reactions with production of slow particles},
  \emph{Sov. Phys. JETP}  \textbf{1}, 2 (1955).

 
\bibitem{Sjostrand:2007gs} 
  T.~Sjostrand, S.~Mrenna and P.~Z.~Skands,
  \emph{A Brief Introduction to PYTHIA 8.1},
  \emph{Comput. Phys. Commun.}  \textbf{178}, 852 (2008)
  [arXiv:0710.3820 [hep-ph]].

\bibitem{Bahr:2008pv} 
  M.~Bahr, S.~Gieseke, M.~A.~Gigg, D.~Grellscheid, K.~Hamilton, O.~Latunde-Dada, S.~Platzer and P.~Richardson {\it et al.},
  \emph{Herwig++ Physics and Manual},
  \emph{Eur. Phys. J.} \textbf{C 58}, 639 (2008)
  [arXiv:0803.0883 [hep-ph]].
  
\bibitem{Buckley:2010ar} 
  A.~Buckley, J.~Butterworth, L.~Lonnblad, H.~Hoeth, J.~Monk, H.~Schulz, J.~E.~von Seggern and F.~Siegert {\it et al.},
  \emph{Rivet user manual},
  arXiv:1003.0694 [hep-ph].

\bibitem{ATLAS:luminosity}
\url{
https://twiki.cern.ch/twiki/bin/view/AtlasPublic/LuminosityPublicResults } .

\bibitem{CMS:luminosity}
\url{https://twiki.cern.ch/twiki/bin/view/CMSPublic/LumiPublicResults}.

\bibitem{Aubert:2009ah} 
  B.~Aubert {\it et al.}  [BaBar Collaboration],
   \emph{Study of $D_{sJ}$ decays to $D^{*}K$ in inclusive $e^{+}e^{-}$ interactions},
  Phys.\ Rev.\ D {\bf 80}, 092003 (2009)
  [arXiv:0908.0806 [hep-ex]].

\bibitem{ATLAS:2011fea} 
  ATLAS Collaboration,
 \emph{Measurement of $D^(*)$ meson production cross sections in pp collisions at $\sqrt{s}=7$ TeV with the ATLAS detector},
  ATLAS-CONF-2011-017.

\bibitem{Roca:2005nm} 
  L.~Roca, E.~Oset and J.~Singh,
  \emph{Low lying axial-vector mesons as dynamically generated resonances},
  \emph{Phys.\ Rev.} D {\bf 72}, 014002 (2005)
  [hep-ph/0503273].



\end{thebibliography}
\end{document}